\documentstyle[aps,preprint]{revtex}
\begin{document}
\tightenlines
\title{Perturbations of the metric induced by 
back - reaction in the warm inflation
scenario}
\author{Mauricio Bellini\footnote{E-mail address: mbellini@mdp.edu.ar}}
\address{Departamento de F\'{\i}sica, Facultad de Ciencias Exactas  y
Naturales \\ Universidad Nacional de Mar del Plata, \\
Funes 3350, (7600) Mar del Plata, Buenos Aires, Argentina.}
\maketitle
\begin{abstract}
A second - order expansion for the 
quantum fluctuations of the matter field
was considered in the framework of the warm inflation scenario.
The friction and Hubble parameters were expanded by means of a
semiclassical approach. 
The fluctuations of the Hubble parameter generates fluctuations
of the metric.
These metric fluctuations produce an effective term of curvature. 
The power spectrum for the metric fluctuations can be calculated
on the infrared sector. 
\end{abstract}
\vskip 2cm
\noindent
PACS number(s): 98.80.Cq, 05.40.+j, 98.70.Vc
\vskip 2cm
\section{Introduction}
Since its early developments, quantum fluctuations has played
an important role in the inflationary 
universe\cite{1,2,lind,2a,linde94,lindex2}.
They lead to cosmological density perturbations that may be
responsible for the origin of structures in the universe\cite{a}
and may completely alter our concepts about the past, the future, and 
global structure of spacetime\cite{b}. A promising approach towards
a better understanding of these phenomena is the paradigm of 
stochastic inflation\cite{c}. 
The aim of stochastic inflation is to
include the quantum contributions in an effective classical theory,
where they appear as stochastic noises.

A natural consequence of this approach is the self - reproduction
of universes and the return to a global
stationary picture. 
The period in which $\ddot a >0$ and thus the universe has an accelerated
expansion is the inflationary stage, and models are discarded or not
depending on the fact that they provide enough inflation or not.
The standard inflationary model separates
expansion and reheating as two distinguished time periods.
This theory assumes an exponential expansion in a second - order phase 
transition of the inflaton field\cite{goncharov}, followed 
by localized mechanism
that rapidly distributes the vacuum energy into thermal energy.
Reheating after inflation occurs due to particle production
by the oscillating inflaton field\cite{lin}.
The differential microwave radiometer (DMR) on the Cosmic Background
Explorer (COBE) has made the first direct probe of the initial density
perturbations through detection of the temperature anisotropies in the 
cosmic background radiation (CBR). The results are consistent with the
scaling spectrum given by the inflation model.
For inflation the simplest assumption is that there are two scales: a 
long - time, long - distance scale associated with vacuum energy dynamics
and a single short - time, short - distance scale associated 
with a random force component. The Hubble time during inflation, $1/H$,
appropriately separates the two regimes. For the grand unified theory\cite{cw},
this time interval is $1/H \sim 10^{-34}$ sec.
Inflation predicts that the initial density perturbations should be
Gaussian, with a power - law spectrum index $n \sim 1$\cite{2b}.
Furthermore, the existence of nucleation of matter in the universe
should be a consequence of these early perturbations of the matter field.

The warm inflation scenario takes into account separately, 
the matter and radiation energy fluctuations. 
This formalism, that
mixes these two 
isolated exponential expansion and reheating stages, may solve the
disparities created by the two each periods. In this 
scenario introduced by Berera and Fang\cite{3,3a},
the thermal fluctuations could play the dominant role
in producing the initial perturbations.
The warm inflation scenario served as a explicit demonstration 
that inflation can occur in the
presence of a thermal component.

In an alternative formalism for warm inflation\cite{4,5} 
I demonstrate that, for a potential inflation
model both, thermal equilibrium and quantum to classical transition
of the coarse - grained field hold for a 
sufficiently large rate of expansion
of the scale factor of the universe. 
For a power - law expanding for scale factor,
the mean temperature decreases
as $t^{-1/2}$.
So, one can predicts
the dynamics for the mean temperature and the amplitude of their
fluctuations.
Thus, at the
end of the inflationary era the thermal equilibrium holds and 
the spectrum of the coarse - grained
matter field can be calculated. 

Any local energy perturbations during inflation can affect a region
of characteristic physical length $1/H(\varphi)$ or less. 
The largest scales of energy density fluctuations
in the post - inflationary universe arose from the earliest perturbations
during inflation. For inflation we understand naturality as both macroscopic
and microscopic. Macroscopically, we would like a description that rests
with common - day experience. Microscopically, it should be consistent
with the standard model of particle physics.
The main drawback of these approaches (i.e., the standard and 
warm inflation ones) is the
slow - roll assumption itself, which gives the reduction
of the equation of motion of the scalar field to a first - order one.

In this work
I develope a formalism
where the quantum field of matter $\varphi$
interacts with other fields of a thermal bath at
mean temperature $<T_r> \  < T_{GUT} \sim 10^{15}$ GeV. 
This lower temperature condition implies that magnetic monopole
suppression works effectively.
In this model
quantum fluctuations of the matter field $\varphi$
lead to quantum fluctuations of the metric
in addition to matter and radiation
energy densities. 
I consider a general case, where the inflaton field has a nonzero
mean value and I develope the analysis by using a consistent
semiclassical expansion.
This work is organized as follow:
In section II) I develope the formalism for a quantum perturbed
flat FRW metric.
Classical and quantum dynamics are studied with a semiclassical approach
on a flat FRW background metric given by the expectation value
of the perturbed flat FRW metric.
In section III) I study the coarse - grained approach for the 
matter and metric
fluctuations to
warm inflation. The expression for the power spectrum of the
metric fluctuations is obtained.
Finally, in section IV) some final remarks are developed.

\section{Formalism}

In the warm inflation
era, the radiation 
energy density must be small with respect to the vacuum energy,
which is given by the potential energy density
$V(\varphi)$. Furthermore
the kinetic component of the energy density is negligible with respect
to the vacuum energy density
\begin{displaymath}
\rho(\varphi) \sim \rho_m \sim V(\varphi) \gg \rho_{kinetic}.
\end{displaymath}
where $\rho_{kinetic} = \rho_r(\varphi)+ {1\over 2} \dot\varphi^2$
and
$\rho_r(\varphi) = {\tau(\varphi)\over 8H(\varphi)} \dot\varphi^2$.
Here, $H(\varphi)$ and $\tau(\varphi)$ are the Hubble and friction parameters.
The conventional treatment for the scalar field dynamics assumes
that it is pure vacuum energy dominated. The various kinematic
outcomes are a result of specially
chosen Lagrangians. In most cases the Lagrangian is unmotivated
from particle phenomenology. Clear exceptions are the 
Coleman - Weinberg potential with a coupling
constant, which is motivated by grand unified theories and
supersymmetric potentials\cite{cw}. Making an
extension to the new inflation picture, the
behavior of the scale factor can also be altered for
any given potential when radiation energy is present.
In our case the density Lagrangian 
that describes the warm inflation scenario is
\begin{equation}
{\cal L}(\varphi,\varphi_{,\mu}) = - \sqrt{-g}\left[\frac{R}{16 \pi}
+\frac{1}{2} g^{\mu \nu} \varphi_{,\mu} \varphi_{,\nu}
+V(\varphi) \right]+{\cal L}_{int}.
\end{equation}
Here,  $R$ is the scalar curvature, $g^{\mu\nu}$ the metric
tensor, $g$ is the metric. The Lagrangian ${\cal L}_{int}$ 
takes into account
the interaction of the field $\varphi$ with other particles
in the thermal bath.
All particlelike matter which existed before
inflation would have been dispersed by inflation.

As in previous works\cite{4,5,6,7}, I consider a semiclassical 
approach for a quantum operator $\varphi(\vec x,t)$
\begin{equation}\label{u}
\varphi(\vec x,t)=\phi_c(t)+\phi(\vec x,t),
\end{equation}
where $<E|\varphi|E>=\phi_c(t)$ is the expectation value of the operator
$\varphi$ in an arbitrary state $|E>$.
Furthermore, I require that $<E|\dot\phi|E>=0$  and 
$<E|\phi(\vec x,t)|E>=0$. 

One can write the perturbed Hubble parameter as an expansion in $\phi$
\begin{equation}\label{b}
H(\varphi)= H_c(\phi_c)+ \sum^{\infty}_{n=1} 
\frac{1}{n!} H^{(n)}(\phi_c) \phi^n,
\end{equation}
where $H_c(\phi_c) \equiv H(\phi_c)$ and $H^{(n)}(\phi_c)\equiv
\left.{d^n H(\varphi) \over d\varphi^n}\right|_{\phi_c}$. I will
consider the quantum fluctuations $\phi$ as very small. 
Thus, will be sufficient to consider a $\phi$ -  first - order expansion
in eq. (\ref{b}).

Now we consider a perturbed flat Friedmann - Robertson - Walker (FRW)
metric
\begin{equation}
ds^2 = - dt^2 + a^2(t) \left[ 1 + h(\vec x,t) \right] d\vec x^2,
\end{equation}
where $a(t) = a_o \  e^{\int H_c(\phi_c) \  dt}$ is the classical
scale factor of the universe and $h(\vec x,t)$ represents the
quantum fluctuations of the metric, such that
\begin{equation}
1+ h(\vec x,t) = e^{2 \int H'(\phi_c) \phi(\vec x,t) \  dt}.
\end{equation}
When the fluctuations $\phi(\vec x,t)$ are very small the field
$h(\vec x,t)$ can be approximated to
\begin{equation}\label{kiii}
h(\vec x,t) \simeq 2 \int H'(\phi_c) \  \phi(\vec x,t) \  dt.
\end{equation}
Note that $<E| h(\vec x,t)|E> =0$ and thus
\begin{equation}
<E| ds^2|E> = - dt^2 + a^2(t) \  d\vec x^2,
\end{equation}
which gives a globally flat FRW metric.
Then, the expectation value of the metric $<E|ds^2|E>$ gives
the background metric.

The quantum equation of motion for the operator $\varphi$ in a globally
flat FRW spacetime is
\begin{equation}\label{as}
\ddot\varphi - \frac{1}{a^2(t)}
\nabla^2 \varphi + \left[3 H(\varphi)+
\tau(\varphi)\right] \  \dot\varphi + V'(\varphi)=0.
\end{equation}
The expression (\ref{as}) with $\tau(\varphi) = 0$ gives the
equation of motion for standard inflation\cite{7}. 
In this work
I will study the case where $\tau(\varphi) \neq 0$.
As the inflation relaxes toward
its minimum energy configuration, it will decay into lighter fields,
generating an effective viscosity\cite{BGR}. 
If this viscosity is large enough,
the inflaton will reach a slow - roll regime, where its dynamics
becomes overdamped. This overdamped regime has been analyzed 
in Ref.\cite{BGR1}.

The semiclassical Friedmann equation for a globally flat FRW metric is
\begin{equation}\label{as1}
\left<E \left| H^2(\varphi)\right|E\right>
=\frac{8 \pi}{3 M^2_p} \left<E\left| \rho_m+\rho_r \right|E\right>,
\end{equation}
where $M_p = 1.2 \  10^{19}$  GeV is the Planckian mass.
Here, the matter and radiation energy densities,
$\rho_m(\varphi) $ and $\rho_r(\varphi)$, are
\begin{eqnarray}
\rho_m(\varphi) &=& \frac{\dot\varphi^2}{2} + \frac{1}{2 a^2}
\left(\vec\nabla \varphi \right)^2+ V(\varphi), \\
\rho_r(\varphi) &=& \frac{\tau(\varphi)}{8 H(\varphi)} \dot\varphi^2.
\end{eqnarray}
In the limiting case of standard inflation [$\tau(\varphi) =0$], the
radiation energy density becomes zero.

We can write $V'(\varphi)$ and the friction parameter $\tau(\varphi)$
in eq. (\ref{as}), as $\phi$ - expansions
\begin{eqnarray}
V'(\varphi) & = & V'(\phi_c)+ \sum^{\infty}_{n=1} \frac{1}{n!}
V^{(n+1)} \phi^{n}, \label{b1} \\
\tau(\varphi) & = & \tau_c + \sum^{\infty}_{n=1} \frac{1}{n!}
\tau^{(n)}(\phi_c) \phi^n. \label{bb1}
\end{eqnarray}

Replacing $V'(\varphi)$ (at second order in $\phi$) with
$H(\varphi)$, $\tau(\varphi)$ and $\varphi$ (at first order
in $\phi$) in eqs. (\ref{as}) and (\ref{as1}),
we obtain the following motion and Friedmann equations 
\begin{eqnarray}
\ddot\phi &+& \ddot\phi_c-\frac{1}{a^2(t)}
\nabla^2(\phi_c+\phi)+
\left[3(H_c+H'\phi)+(\tau_c+\tau'\phi)\right]
(\dot\phi_c+\dot\phi)  \nonumber \\
&+& V'(\phi_c)+ V''(\phi_c) \phi+\frac{1}{2}
V'''(\phi_c) \phi^2=0,\label{y2}
\end{eqnarray}
and
\begin{eqnarray}
\left<E\left|\left(H_c+H' \phi\right)^2\right|E\right> &=& 
\frac{4\pi}{3M^2_p}\left<E\left|
(\dot\phi_c+\dot\phi)^2 \left[1+\frac{\tau_c+\tau'\phi}{4(H_c
+H'\phi)}\right] +\frac{1}{a^2(t)}
\left[\vec\nabla(\phi_c+\phi)\right]^2\right.\right. \nonumber \\
& + &\left.\left.
2\left[V(\phi_c)+V'(\phi_c)\phi+\frac{1}{2} V''(\phi_c) \phi^2\right]
\right|E \right>,\label{y3}
\end{eqnarray}
where we have expanded the 
Hubble and friction parameters [$H(\varphi)$ and $\tau(\varphi)$],
at first order in $\phi$.
Furthermore, $V(\varphi)$ and $V'(\varphi)$ were expanded 
at second order in $\phi$. 
The friction
parameter takes into account the interaction of the matter field
with the fields in the thermal bath. 
The eq. (\ref{y2}) describes the dynamics for a semiclassical expansion
of the matter field $\varphi$. The eq. (\ref{y3}) is the second order 
semiclassical approach for the
Friedmann equation with this expansion for $\varphi$.

\subsection{Dynamics of the classical field}

We consider the eq. (\ref{y2}) at zero order in $\phi$. The equation of motion
for the field $\phi_c(t)$ is
\begin{equation}\label{bu}
\ddot\phi_c+ [3H_c(\phi_c)+\tau_c(\phi_c)] \dot\phi_c+ V'(\phi_c)=0.
\end{equation}
Note that making $\tau_c= 0$,
$|\ddot\phi_c| \ll 3 H_c(\phi_c) \  |\dot\phi_c|$ and
$|\ddot \phi_c| \ll V'(\phi_c)$ one obtains the limit
case for slow - roll regime in standard inflation.
Furthermore, for $\tau_c(\phi_c) \gg 3 H_c(\phi_c)$ (i.e., for
$\gamma \gg 1$), $|\ddot\phi_c| \ll \tau_c(\phi_c) |\dot\phi_c|$ and
$|\ddot\phi_c| \ll V"(\phi_c)$, one recover the slow - roll limiting regime
in warm inflation.
The dynamics for both, the classical field $\phi_c$
and the classical Hubble parameter $H_c(\phi_c)$, are
\begin{eqnarray}
\dot\phi_c &=& - \frac{M^2_p}{4\pi}H'_c 
\left(1+ \frac{\tau_c}{3H_c}
\right)^{-1}, \label{bp}\\
\dot H_c & =& H'_c \dot\phi_c = - \frac{M^2_p}{4\pi} (H'_c )^2
\left(1+ \frac{\tau_c}{3 H_c}\right)^{-1}. \label{bp1}
\end{eqnarray}
Observe that $\dot H_c <0$, which means that the classical Hubble
parameter decreases with time. Equations (\ref{bp}) and (\ref{bp1})
define the classical evolution of the spacetime (i.e., the background
spacetime).
The eq. (\ref{y3}) at zero order in $\phi$, gives
the classical Friedmann equation 
\begin{equation}\label{a11}
H^2_c(\phi_c)=\frac{4\pi}{3 M^2_p}\left[ \left(1+
\frac{\tau_c}{4 H_c}\right) \dot\phi^2_c+ 2 V(\phi_c)\right].
\end{equation}
From eqs. (\ref{bp}) and (\ref{a11}), the classical potential becomes  
\begin{equation}\label{ap}
V(\phi_c) = \frac{3 M^2_p}{8\pi}\left[ H^2_c(\phi_c)-
\frac{M^2_p}{12\pi}\left(H'_c\right)^2 \left(1+\frac{\tau_c}{4 H_c}
\right)
\left(1+\frac{\tau_c}{3 H_c}\right)^{-2}\right].
\end{equation}
The effective classical potential depends on both, the friction and
Hubble parameters.
The classical expressions the matter and
radiation energy densities, are
\begin{eqnarray}
\rho_m(\phi_c)  &=& \frac{1}{2}\left[\dot\phi^2_c
+ 2 V(\phi_c)\right], \label{q}\\
\rho_r(\phi_c)  &=&  
\frac{\tau_c}{8 H_c} \  \dot\phi^2_c . \label{q1}
\end{eqnarray}
The characteristic time scale for inflation is $\tau_H = 1/H_c$, and
the number of e - folds in inflation is
\begin{equation}
N_e = \int^{t_{end}}_{t_o} \frac{dt}{\tau_H}.
\end{equation}

The basic idea of warm inflation is quite simple. A scalar field
coupled to several fields in a thermal bath. As the inflaton relaxes
toward its minimum energy configuration, it will decay into lighter
fields, generating an effective friction. If this friction is
large enough, the inflation will reach a slow - roll regime, where
its dynamics becomes overdamped. In order to satisfy one of the
requirements of a successful inflation (60 or more e-folds),
overdamping must be very efficient\cite{3a}.

In the next subsections I will develope the dynamics for the quantum
field $\phi$ which gives information about the spatial inhomogeneities
of matter in the universe. 

\subsection{first - order Dynamics of the quantum perturbations}

From eq. (\ref{y2}) one obtains
the equation of motion for
a $\phi$ - first - order expansion
\begin{equation}\label{dd}
\ddot\phi - \frac{1}{a^2(t)} \nabla^2\phi 
+ 3 H_c\left(1+\frac{\tau_c}{3H_c}\right) \dot\phi+
\left[3H'\left(1+\frac{\tau'}{3H'}\right)\dot\phi_c + V''\right]\phi = 0.
\end{equation}
The Friedmann equation becomes [see eq. (\ref{y3})]
\begin{equation}\label{zi}
\left<E\left| 2 H_c H' \phi \right|E\right> = \frac{4\pi}{3 M^2_p} 
\left<E\left|
\dot\phi_c \left[\frac{\dot\phi_c}{4H_c} \left[\tau'-\frac{H'\tau_c}{4H_c}\right]
\phi
+2\left(1+\frac{\tau_c}{4H_c}\right)\dot\phi\right] + 2 V'(\phi_c) \phi \right|E\right>,
\end{equation}
which is zero, due to $<E|\phi(\vec x,t)|E>=<E|\dot\phi(\vec x,t)|E>=0$.
Furthemore, from eq. (\ref{bp}) one obtains the equation of motion for the
matter field fluctuations $\phi$
\begin{equation}\label{zie}
\ddot\phi - \frac{1}{a^2(t)} \nabla^2 \phi +
\left[3H_c + \tau_c\right] \dot\phi +
\left[V''(\phi_c) - \frac{3 M^2_p}{4 \pi}
\left(H'\right)^2 \left(1+\frac{\tau'}{3H'}\right)\left(1+
\frac{\tau_c}{3H_c}\right)^{-1}\right] \phi = 0,
\end{equation}
where I have used $H'H'_c = (H')^2$.

We can introduce the new field 
$\chi = e^{3/2 \int (H_c+\tau_c/3) dt} \  \phi$
to describe the quantum fluctuations of the matter field. 
The eq. (\ref{zie}), written as a function
of $\chi$, is
\begin{equation}\label{ij}
\ddot\chi - \frac{1}{a^2} \nabla^2 \chi - \frac{k^2_o(t)}{a^2(t)} \chi = 0,
\end{equation}
with
\begin{equation}\label{ero}
k^2_o(t) =  a^2(t) 
\left[\frac{3 M^2_p}{4\pi} \left(H'\right)^2
\left(1+\frac{\tau_c}{3 H_c}\right)^{-1}\left(1+\frac{\tau'}{3H'}\right)
+\frac{9}{4} \left(H_c+\frac{\tau_c}{3}\right)^2+
\frac{3}{2}\left(\dot H_c + \frac{\dot\tau_c}{3}\right)
-V''(\phi_c) \right]
\end{equation}
The eq. (\ref{ij}) is a Klein - Gordon equation for $\chi$ in a
globally flat FRW background metric, with a time - dependent 
parameter of mass $\mu(t) = {k_o \over a}$.
Note that $k_o$ depends with $t$ [see eq. (\ref{ero})].
The very important difference between $k_o$ in eq. (\ref{ero})
and other approximations is that here the appears a new
term
$\left(1+{\tau_c\over 3 H_c}\right)^{-1}{\tau'\over3H'}$
due to the fact that we have considering
the fluctuations for the Hubble and friction parameters.

Now we define the quantum perturbations $\chi$, as a Fourier
expansion in terms of the modes $\xi_k(t) e^{i \vec k. \vec x}$
\begin{equation}
\chi(\vec x,t) = \frac{1}{(2\pi)^{3/2}} \int d^3 k
\left[ a_k \xi_k(t) e^{i \vec k. \vec x} + H.c\right],
\end{equation}
where $a^{\dagger}_k$ and $a_k$ are the creation and annihilation
operators
\begin{equation}
[a_k, a^{\dagger}_{k'}] = \delta^{(3)} (\vec k - \vec k'); \qquad
[a_k, a_{k'}] = [a^{\dagger}_k, a^{\dagger}_{k'}] =0.
\end{equation}
From eq. (\ref{kiii}), the field $h(\vec x,t)$ becomes
\begin{equation}
h(\vec x, t) = \frac{1}{(2\pi)^{3/2}} \int d^3 k
\left[ a_k {\mathaccent "707E\xi}_k(t) e^{i \vec k. \vec x} + H.c\right],
\end{equation}
which is the field that describes the quantum fluctuations of the metric
$ds^2$, written as a Fourier expansion in terms of the modes
${\mathaccent "707E\xi}_k(t) e^{i \vec k. \vec x}$.
The time - dependent modes ${\mathaccent "707E\xi}_k(t)$ are
\begin{equation}\label{34}
{\mathaccent "707E\xi}_k(t) = 2 
\int H'(t) \ \xi_k(t) e^{-3/2\int^{t} (H_c+\tau_c/3) \  dt'} \  dt.
\end{equation}
The commutation relation between $\chi$ and $\dot\chi$ is
\begin{equation}\label{ki}
[\chi(\vec x,t), \dot\chi(\vec x',t)] = 
{\rm i} \  \delta^{(3)} (\vec x -\vec x'),
\end{equation}
which is valid for the following relation of the time - dependent modes
$\xi_k(t)$:
\begin{equation}
\xi_k \dot\xi^*_k - \dot\xi_k \xi^*_k ={\rm i}.
\end{equation}
Hence, the commutation relation
between $h(\vec x,t)$ and $\dot h(\vec x,t)$ is
\begin{equation}\label{ko}
[h(\vec x,t), \dot h(\vec x,t)] = \frac{1}{(2\pi)^3}
\int d^3 k \  ({\mathaccent "707E\xi}_k \dot{\mathaccent "707E\xi}^*_k
- \dot{\mathaccent "707E\xi}_k
{\mathaccent "707E\xi}^*_k ) \  e^{-i \vec k. (\vec x - \vec x')},
\end{equation}
which can be calculated once one obtains the modes $\xi_k$ and the Hubble
parameter $H_c(t)={\dot a(t) \over a(t)}$.
Furthermore, the equation of motion for the time - dependent 
modes $\xi_k(t)$ is
\begin{equation}\label{zt}
\ddot\xi_k(t) + a^{-2}(t) \left[ k^2 - k^2_o(t)\right] \xi_k(t) = 0,
\end{equation}
for $k_o(t)$ given by eq. (\ref{ero}). Equation (\ref{zt}) is the
same that the equation of an harmonic 
oscillator with square time - dependent
frequency $\omega^2_k = a^{-2}(t)\left[k^2 - k^2_o(t)\right]$.

\subsection{second - order perturbations for the matter field}

The dynamics of the quantum fluctuations for the matter field at
first - order in $\phi$ was studied with detail in the last subsection. 
In this
subsection I will develope the basic statments for
perturbations of the matter field at second - order in $\phi$.

Equating the terms at second - order in $\phi$ in eq. (\ref{y2}), one 
obtains
\begin{equation}\label{us} 
\left(3 H'+\tau'\right) \phi \dot\phi + 1/2 V''' \phi^2=0, 
\end{equation}
and the Friedmann eq. (\ref{y3}), at second - order in $\phi$, is
\begin{eqnarray}
&& \left< E \left|(H')^2 \phi^2 \right|E \right> \simeq  
\frac{4 \pi}{3 M^2_p}\left<
E \left| \dot\phi^2 \left(1+\frac{\tau_c}{4 H_c}\right)
+\frac{1}{a^2(t)} \left[\vec\nabla\phi\right]^2\right.\right. \nonumber \\
&+ &
\left\{ V''(\phi_c) 
+\frac{M^2_p}{4\pi} H'_c
\left(1+\frac{\tau_c}{3H_c}\right)^{-1} \right. \nonumber \\
&\times & \left.\left.\left. \left[
\frac{1}{4H_c} \left(\tau' - \frac{H'\tau_c}{4 H_c}\right)
\frac{V'''}{(3H'+\tau')} -
\frac{M^2_p}{4\pi} H'_c \left(1+\frac{\tau_c}{3H_c}\right)^{-1}
\frac{\tau'H'}{(4H_c)^2}\right]\right\}
\phi^2 \right|E\right>, \label{usf}
\end{eqnarray}
where the approximation ${1 \over 4 H_c[1+H'\phi/(4H_c)]} \simeq
{1\over 4H_c} \left(1- {H'\phi \over 4 H_c}\right)$ was used. Due to
$\dot h(\vec x,t) = 2 H' \phi(\vec x,t)$,
the eq. (\ref{usf}) can be written as
\begin{eqnarray}
&& \left< E \left| \frac{(\dot h)^2}{4}\right|E\right>  \simeq 
\frac{4\pi}{3 M^2_p} \left<E\left| \dot\phi^2 \left(
1+ \frac{\tau_c}{4 H_c}\right) 
+  \frac{1}{a^2} \left(\vec\nabla \phi\right)^2 \right.\right. \nonumber \\
&+& 
\left\{ V''(\phi_c) 
+\frac{M^2_p}{4\pi} H'_c
\left(1+\frac{\tau_c}{3H_c}\right)^{-1}\right. \nonumber \\
& \times & \left.\left. \left. \left[
\frac{1}{4H_c} \left(\tau' - \frac{H'\tau_c}{4 H_c}\right)
\frac{V'''}{(3H'+\tau')} -
\frac{M^2_p}{4\pi} H'_c \left(1+\frac{\tau_c}{3H_c}\right)^{-1}
\frac{\tau'H'}{(4H_c)^2}\right]\right\}
\phi^2 \right|E\right>,
\label{42a}
\end{eqnarray}
which gives the effective curvature 
of the spacetime due to the fluctuations
of the matter field. Thus, the term $\left<E\left| {(\dot h)^2\over 4}
\right|E\right> = {K \over a^2}$ gives an additional contribution in the
semiclassical Friedmann equation, such that the eq. (\ref{y3}) becomes
\begin{eqnarray}
&& H^2_c(\phi_c) + \frac{K}{a^2} = 
\frac{8\pi}{3 M^2_p} V(\phi_c) + \frac{M^2_p}{12 \pi}
\left(H'\right)^2 \left(1+\frac{\tau_c}{4 H_c}\right)
\left(1+\frac{\tau_c}{3 H_c}
\right)^{-2} \nonumber \\
&+& \frac{8\pi}{3M^2_p} \left< E \left| \dot\phi^2 \left(
1+\frac{\tau_c}{4 H_c}
\right)+ \frac{1}{a^2} \left(\vec\nabla \phi\right)^2 
+ 
\left\{ V''(\phi_c) 
+\frac{M^2_p}{4\pi} H'_c
\left(1+\frac{\tau_c}{3H_c}\right)^{-1} \right.\right.\right. \nonumber \\
& \times & \left.\left.\left.\left[
\frac{1}{4H_c} \left(\tau' - \frac{H'\tau_c}{4 H_c}\right)
\frac{V'''}{(3H'+\tau')} -
\frac{M^2_p}{4\pi} H'_c \left(1+\frac{\tau_c}{3H_c}\right)^{-1}
\frac{\tau'H'}{(4H_c)^2}\right]\right\}
\phi^2 
\right| E \right>.\label{42}
\end{eqnarray}
Hence, the second - order fluctuations of the matter field introduces an additional
curvature in the background spacetime $<E|ds^2|E>=-dt^2+a^2(t) d\vec x^2$.
The expression (\ref{42}) can be written as
\begin{equation}
H^2_c(\phi_c) + \frac{K}{a^2} = \frac{8\pi}{3 M^2_p}
\left< E \left| \rho_m + \rho_r \right|E\right>.
\end{equation}
Here, $<E|\rho_m|E>$ and $<E|\rho_r|E>$, are the effective matter and
radiation energy densities, which include the second - order fluctuations
of the matter field. The mean temperature of the bath is given by
\begin{equation}\label{46}
<T_r> \propto \left[<E|\rho_r|E>\right]^{1/4}.
\end{equation}

\section{The coarse - grained fields}

To develope a stochastic treatment for quantum fluctuations of the 
matter field we must study the universe on a scale much
greater than
the scale of the observable universe. Thus, I consider the 
redefined quantum fluctuations $\chi$ as two pieces
\begin{equation}
\chi=\chi_{cg} + \chi_S. 
\end{equation}
The piece $\chi_{cg}$
takes into account only 
the modes with wavelength of size
\begin{displaymath}
l \ge \frac{1}{\epsilon k_o},
\end{displaymath}
which is much bigger than the size of the horizon (with size 
$1/k_o$).
This coarse - grained field is given by
\begin{equation}\label{t}
\chi_{cg}(\vec x,t)= \frac{1}{(2\pi)^{3/2}} \int
d^3k \  \theta(\epsilon k_o - k) \  \left[ a_k e^{i \vec k.\vec x}
\xi_k + H.c.\right],
\end{equation}
where   $\theta(\epsilon k_o - k)$ is a Heaviside function which
acts as a suppression factor, and
$\epsilon \ll 1$ is a constant. Moreover, one can define the 
coarse - grained  field that represent the fluctuations of the metric on the
infrared sector
\begin{equation}
h_{cg}(\vec x,t) = \frac{1}{(2\pi)^{3/2}} \int d^3k \  \theta
(\epsilon k_o -k) \  \left[ a_k e^{i \vec k. \vec x} 
{\mathaccent "707E\xi}_k(t) + H.c. \right],
\end{equation}
where $h = h_{cg} + h_S$.
On the other hand, the pieces $\chi_S$ and $h_S$ take into account
the modes with wanumbers greater than $\epsilon k_o$ 
\begin{eqnarray}
\chi_{S}(\vec x,t) &=& \frac{1}{(2\pi)^{3/2}} \int
d^3k \  \theta( k-\epsilon k_o ) \  \left[ a_k e^{i \vec k.\vec x}
\xi_k + H.c.\right],                  \\
h_{S}(\vec x,t) &=& \frac{1}{(2\pi)^{3/2}} \int
d^3k \  \theta( k-\epsilon k_o ) \  \left[ a_k e^{i \vec k.\vec x}
{\mathaccent "707E\xi}_k(t) + H.c.\right].
\end{eqnarray}
The quantum fluctuations with wavenumbers greater than $\epsilon k_o$ are
generally interpreted as inhomogeneities superimposed to the
classical field. They are responsible for the density
inhomogeneities generated during the inflation.
The modes with $k/k_o > \epsilon$ are refered to as outside
the superhorizon.

\subsection{Quantum to Classical transition for the coarse - grained field}

Quantum to classical transition of the quantum fluctuations $\phi$
occurs when the commutators (\ref{ki}) and (\ref{ko}) 
are zero. 
When this transition holds one obtains
$\xi_k \dot\xi^*_k - \dot\xi_k \xi^*_k \simeq 0$.
We can represent the modes by
\begin{equation}
\xi_k(t)= u_k(t)+ {\rm i} \  v_k(t),
\end{equation}
where $u_k(t)$ and $v_k(t)$ are time - dependent real functions.
The condition to obtain the complex to real transition of a given
mode $\xi_k$ is
\begin{equation}\label{60}
\left|\frac{v_k(t)}{u_k(t)}\right| \ll 1.
\end{equation}
The quantum to classical transition function (QCTF)
$\alpha_k(t)=\left|{v_k(t)\over u_k(t)}\right|$ was defined in previous
works\cite{5}.
The modes are real when this function becomes nearly zero.
The condition for the coarse-grained fields, $\chi_{cg}(\vec{x},t)$
and $h_{cg}(\vec{x},t)$,
become classical when the modes with $k< \epsilon k_o$
become real. More exactly
\begin{equation}
\frac{1}{N(t)}\sum_{k=0}^{k= \epsilon k_o}  \alpha_k(t) \ll 1,
\end{equation}
where $N(t)$ is time - dependent
number of degrees of freedom of the infrared sector.
During inflation $N(t)$ increases with time, since
$k_o(t)$ is also increasing with time. 
Hence, constantly new degrees of freedom
with a given $k \ll k_o$ enters in the infrared sector.
Quantum to classical
transition for the coarse-grained fields, $\chi_{cg}$ and
$h_{cg}$, occurs when
$\alpha_{k=\epsilon k_o}(t)\ll 1$.
It can be seen that the solutions of eq. (\ref{zt})
for $k < k_o$ are real, while for $k > k_o$ the time - dependent modes
are complex.
Hence, when (\ref{60}) is satisfied by all the modes which were much
bigger than the size of the horizon,
the time - dependent modes $\xi_k$ become real
in the infrared sector. This condition impose restrictions on the vacuum.
A similar condition than (\ref{60}) also was obtained in the framework
of standard inflation by D. Polarski and A. A. Starobinsky\cite{CQG1,CQG2},
but for $\alpha_{k} \gg 1$. However, both conditions (\ref{60})
and the Polarski - Starobinsky's one, are equivalent.

\subsection{Coarse - grained field fluctuations}

When the quantum fluctuations become classical, the amplitude of the
fluctuations
of the field $\phi^{(c)}_{cg}$ are (in the following I 
denote $<E|...|E>$ as $<...>$)\cite{np}
\begin{equation}
\left<[\phi^{(c)}_{cg}]^2(t)\right> 
= \frac{e^{-3 \int^{t}_{} 
(H_c+\tau_c/3) dt'}}{2 \pi^2 } \int^{\epsilon k_o}_{0} 
dk \  k^2 \  \xi^2_k(t).
\end{equation}
The square fluctuations
\begin{equation}\label{flu}
<[h^{(c)}_{cg}(t)]^2> = \frac{1}{2\pi^2} \int^{\epsilon k_o}_{0}
dk \  k^2 \  {\mathaccent "707E\xi}^2_k(t),
\end{equation}
give the temporal evolution for the amplitude for the fluctuations
in the metric on the infrared sector. Note that $\tilde\xi_k(t)$ depends
with $\tau_c$ [see eq. (\ref{34})].
Thus, the square fluctuations $\left<\left[h^{(c)}_{cg}\right]^2\right>$
are $\tau_c$ - dependent.
To study the power spectrum ${\cal P}_{h^{(c)}_{cg}}(k)$ for the fluctuations
of the metric, one can write the square fluctuations
$\left<\left[h^{(c)}_{cg}\right]^2\right>$ as
\begin{equation}
\left<\left[h^{(c)}_{cg}(t)\right]^2\right> = \frac{1}{2\pi^2} 
\int^{\epsilon k_o(t)}_{0} \frac{dk}{k} {\cal P}_{h^{(c)}_{cg}}(k),
\end{equation}
where the power spectrum for the fluctuations of the metric,
${\cal P}_{h^{(c)}_{cg}}(k)$, when the horizon
entry is 
\begin{equation}\label{79}
{\cal P}_{h^{(c)}_{cg}}(k) = A(t_*) \left(\frac{k}{\epsilon k_o(t_*)}
\right)^n f(k),
\end{equation}
where $t_*$ denotes the time when the horizon entry, for which
$k_o(t_*) \simeq \pi H_o$ in comoving scale.
Furthermore, the factor $A(t_*)$ gives the amplitude of the spectrum
when the horizon entry, $n$ is the spectral index, and $f(k)$ is the
square - $t_*$ - evaluated Heaviside function: 
$\theta^2(\epsilon k_o - k)$.
The standard choice $n=1$ gives a scale invariant
spectrum. However, an experimentally constraint for the COBE data
gives\cite{Bar}
\begin{displaymath}
|n-1.2| < 0.3.
\end{displaymath}

\section{Final Remarks}

Warm inflation takes into account separately, the
matter and radiation energy densities. During the
inflationary era the mean temperature
is smaller than the GUT one (i.e., $<T_r> \  < \  T_{GUT}
\simeq 10^{15}$ GeV).
This lower temperature condition implies that magnetic monopole
suppression works effectively.
In this work I developed a semiclassical formalism for warm
inflation that takes into account the fluctuations of the
metric around a flat FRW background metric.
The fundamental aspects of the formalism here developed are the following:

1) I consider a semiclassical expansion for the field $\varphi = \phi_c +
\phi$. In this framework, the 
Hubble and 
friction parameters were expanded at firsth order in $\phi$.
The semiclassical expansion for the Hubble parameter generates
a second - order semiclassical Friedmann eq. (\ref{y3}). This equation
takes into account the expectation value for both, matter and radiation
energy densities. The radiation energy density 
is proportional to the mean temperature
($<\rho_r> \propto <T_r>^4$) of the thermal bath.

2) The equation of motion (\ref{dd}) is valid for a background metric
which describes a globally flat FRW spacetime. However, this metric
fluctuates due to the back - reaction with the fluctuations of the
matter field. This metric fluctuations are described by the field
$h(\vec x,t)$. The eq. (\ref{kiii}) shows how $h(\vec x,t)$ depends
with $\phi(\vec x,t)$. Due to the second - order $\phi$ - expansion
in the semiclassical Friedmann eq. (\ref{y3}), an effective
curvature $K$ arises in the semiclassical Friedmann equation
[see eq. (\ref{42})]. The effective curvature is related with
the metric fluctuations by eq. (\ref{42a}).

3) The field $\phi(\vec x,t)$ can be written as a Fourier expansion.
For this description, the relevant modes $\xi_k(t)$ describe
the temporal evolution of $\phi$. In this work
I studied these perturbations on a scale much bigger than the size
of the horizon. In the infrared sector, the temporal evolution 
for the square
$h^{(c)}_{cg}$ - fluctuations are described by the modes $\tilde\xi_k$
[see eq. (\ref{flu})].
In the infrared sector 
these modes must be real in order to $[\chi_{cg},\dot\chi_{cg}]=
[h_{cg}, \dot h_{cg}]=0$.
The power $h^{(c)}_{cg}$ - spectrum in the infrared sector is given
by eq. (\ref{79}). The modes $\xi_k$ are the solution
of the eq. (\ref{zt}) and $\tilde\xi_k(t)$ can be calculated by means
of (\ref{34}).
In the equation that describes the
evolution for $\chi$ [see eq. (\ref{ij})], an effective
time - dependent parameter of mass $\mu(t) = {k_o(t) \over a(t)}$ appears.

\end{document}